\documentclass[twoside,twocolumn,english,superscriptaddress,showkeys,showpacs]{revtex4-1}
\usepackage[T1]{fontenc}
\usepackage[latin9]{inputenc}
\usepackage{amssymb}
\usepackage{stmaryrd}
\usepackage{graphicx}
\usepackage{esint}
\usepackage{lineno}

\makeatletter

\@ifundefined{textcolor}{}
{%
 \definecolor{BLACK}{gray}{0}
 \definecolor{WHITE}{gray}{1}
 \definecolor{RED}{rgb}{1,0,0}
 \definecolor{GREEN}{rgb}{0,1,0}
 \definecolor{BLUE}{rgb}{0,0,1}
 \definecolor{CYAN}{cmyk}{1,0,0,0}
 \definecolor{MAGENTA}{cmyk}{0,1,0,0}
 \definecolor{YELLOW}{cmyk}{0,0,1,0}
}

\usepackage{epsfig}

\makeatother

\begin{document}

\title{Limiting neutrino magnetic moments with Borexino Phase-II solar neutrino data}

\author{M.~Agostini}
\affiliation{Gran Sasso Science Institute (INFN), 67100 L'Aquila, Italy}
\author{K.~Altenm\"{u}ller}
\author{S.~Appel}
\affiliation{Physik-Department and Excellence Cluster Universe, Technische Universit\"at  M\"unchen, 85748 Garching, Germany}
\author{V.~Atroshchenko}
\affiliation{National Research Centre Kurchatov Institute, 123182 Moscow, Russia}
\author{Z.~Bagdasarian}
\affiliation{Institut f\"ur Kernphysik, Forschungszentrum J\"ulich, 52425 J\"ulich, Germany}
\author{D.~Basilico}
\author{G.~Bellini}
\affiliation{Dipartimento di Fisica, Universit\`a degli Studi e INFN, 20133 Milano, Italy}
\author{J.~Benziger}
\affiliation{Chemical Engineering Department, Princeton University, Princeton, NJ 08544, USA}
\author{D.~Bick}
\affiliation{Institut f\"ur Experimentalphysik, Universit\"at Hamburg, 22761 Hamburg, Germany}
\author{G.~Bonfini}
\affiliation{INFN Laboratori Nazionali del Gran Sasso, 67010 Assergi (AQ), Italy}
% ...

\collaboration{\bf{The Borexino collaboration}}


\begin{abstract}
A search for the solar neutrino effective magnetic moment has been
performed using data from 1291.5 days exposure during the second
phase of the Borexino experiment. No significant deviations from the
expected shape of the electron recoil spectrum from solar neutrinos
have been found, and a new upper limit on the effective neutrino magnetic
moment of $\mu_{\nu}^{eff}$ $<$ 2.8$\cdot$10$^{-11}$ $\mu_{B}$
at 90\% c.l. has been set using constraints on the sum of the solar
neutrino fluxes implied by the radiochemical gallium experiments.
Using the limit for the effective neutrino moment, new limits for
the magnetic moments of the neutrino flavor states, and for the elements of the neutrino magnetic moments matrix for Dirac and Majorana neutrinos, are derived.
\end{abstract}

\keywords{solar neutrinos, magnetic moment}
\pacs{14.60.S, 96.60.J, 26.65, 13.10}

\maketitle

\section{Introduction}

Neutrinos produced in the Sun are a unique source of information with regards to their physical properties. Besides the study of well-established
neutrino oscillations they can also be used to look for an anomalous magnetic
moment and other electromagnetic properties of neutrinos \cite{Vol86}--\cite{Giunti15}.
The neutrino magnetic moment in the standard electroweak theory (SM), when extended to include neutrino mass, is proportional to the neutrino mass 
\cite{Fuj80}--\cite{Shrock82}:

\begin{equation}
\mu_{\nu}=\frac{3m_{e}G_{F}}{4\pi^{2}\sqrt{2}}m_{\nu}\mu_{B}\;\approx\;3.2\times10^{-19}\left(\frac{m_{\nu}}{1eV}\right)\;\mu_{B},\label{Formula:MuNu}
\end{equation}

where $\mu_{B}=\frac{eh}{4\pi m_{e}}$ is the Bohr magneton,
$m_{e}$ is the electron mass, and $G_{F}$ is the Fermi coupling constant.
The known upper limit on the neutrino masses $m_{\nu}$ leads to $\mu_{\nu}$
less than 10$^{-18}\mu_{B}$, which is roughly eight orders of magnitude
lower than existing experimental limits. The most stringent laboratory
bounds on $\mu_{\nu}$ are obtained by studying $(\nu,e)$ elastic-scattering
of solar neutrinos and reactor anti-neutrinos. The Super-Kamiokande Collaboration
achieved a limit of 3.6$\cdot$10$^{-10}$ $\mu_{B}$ (90$\%$ C.L.) by fitting day/night solar neutrino spectra above 5-MeV.
With additional information from other solar neutrino and KamLAND experiments
a limit of 1.1$\cdot$10$^{-10}$ $\mu_{B}$ (90$\%$ C.L.) was obtained
\cite{Liu04}. The Borexino collaboration reported the best current limit on 
the effective magnetic moment of 5.4$\cdot$10$^{-11}$ $\mu_{B}$ (90$\%$ C.L.) using
the electron recoil spectrum from $^{7}$Be solar neutrinos \cite{Be7-2}.

The best magnetic moment limit from 
reactor anti-neutrinos is 2.9$\cdot$ 10$^{-11}$ $\mu_{B}$ (90$\%$
C.L.) \cite{Bed07}. The most stringent limits on the neutrino magnetic
moment of up to $\thicksim$10$^{-12}\mu_{B}$ come from astrophysical
observations \cite{Raffelt88,MuAstro}. The complete historical record
of limits on the neutrino magnetic moment can be found in \cite{PDG16}.

Though experimental bounds on $\mu_{\nu}$ are far from the value
predicted by the extended SM, in more general models, for example with right-handed
bosons or with an extended sector of scalar particles, the magnetic
moment can be proportional to the mass of charged leptons and can
have values close to the experimental limits reported.
In more general models the proportionality between the neutrino mass and
its magnetic moment doesn't hold.

In this paper, we report a search for neutrino magnetic moments using 1270.6 days of data 
collected during the Borexino Phase-II campaign. Borexino
is the first real-time detector of low energy solar neutrinos, located
in the Gran Sasso National Laboratory, Italy. Borexino detects solar
neutrinos via the elastic scattering of neutrinos off electrons in
liquid scintillator. The scattered recoil electrons are detected
via scintillation light, which carries the energy and position
information. The mass of the scintillator (PC+PPO) is 278 tons. Events are
selected within a fiducial volume (FV) corresponding to approximately 1/4 of the scintillator volume in order to provide an "active shield" against external backgrounds. 
Detailed descriptions of the detector can be found in \cite{Ali02,Brx08}.

Neutrino-electron elastic scattering is the most sensitive test
for a neutrino magnetic moment search. In the SM, the scattering of
a neutrino with a non-zero magnetic moment is determined 
by both a weak interaction and a single-photon exchange term. The
latter changes the helicity of the final neutrino state. 
This means that the amplitudes of the weak and electromagnetic scattering do not interfere,
at least at the level of $\sim m_{\nu}/E_{\nu}$, and the total cross
section is the sum of the two.

Neutrino mixing means that the coupling of the neutrino mass eigenstates
i and j to an electromagnetic field is characterized by a 3x3 matrix
of the magnetic (and electric) dipole moments $\mu_{ij}$. For Majorana
neutrinos the matrix $\mu_{ij}$ is anti-symmetric and only transition
moments are allowed, while for Dirac neutrinos $\mu_{ij}$ is a general
3x3 matrix. The electromagnetic contribution to the $\nu$--$e$ scattering
cross section is proportional to the square of the effective magnetic moment
$\mu_{eff}$:

\begin{equation}
\frac{d\sigma_{EM}}{dT_{e}}(T_{e},E_{\nu})=\pi\; r_{0}^{2}\;\mu_{eff}^{2}\left(\frac{1}{T_{e}}-\frac{1}{E_{\nu}}\right),\label{Formula:EM}
\end{equation}

where $\mu_{eff}$ is measured in $\mu_{B}$ units and depends on the
components of the neutrino moments matrix $\mu_{ij}$, $T_{e}$ is electron recoil energy, and $r_{0}=2.818\times10^{-13}$
cm is the classical electron radius. 

The energy dependence for the magnetic and weak scattering cross sections differ significantly; 
for $T_{e}\ll E_{\nu}$ their ratio is proportional to $1/T_{e}$ and the sensitivity of the experiment
to the magnetic moment strongly depends on the threshold of detection. 
This makes the low energy threshold of Borexino suitable for a neutrino magnetic moment search.

\section{Data selection and analysis}

The data used for the analysis was collected from December~14, 2011
to May~21, 2016 with a live-time of 1291.5~days. Events were selected
following the procedure optimized for the new solar neutrino analysis
\cite{Brx17}: all events within 2~ms of any muon were rejected, 
whilst a dead time of 300~ms was applied after muons crossing the inner detector; 
decays due to radon daughters occurring before
$^{214}$Bi-$^{214}$Po delayed coincidences are vetoed; events
must be reconstructed within the FV defined by the
following conditions: $R\leq$3.021~m and $|Z|\leq$1.67~m
where R is the reconstructed distance to the detector center and Z is the reconstructed vertical 
coordinate (the last condition excludes parts of the detector which observe higher than average event rates). 
The cuts reduce the live-time to 1270.6 days and the total FV exposure corresponds to 263.7~tonne$\cdot y$.

The model function fitted to the data has been restricted to the same
components used in the solar neutrino analysis of the second phase (see
\cite{Brx17}), namely $^{14}$C, $^{85}$Kr, $^{210}$Bi $\beta$-decay shapes, the $\beta^{+}$ spectrum of the cosmogenic
$^{11}$C, the monoenergetic $\alpha$ peak from $^{210}$Po decays, $\gamma$-rays from external sources and
the electron recoil spectra from $^{7}$Be, $pp$, $pep$ and the CNO
cycle neutrinos. Other backgrounds and solar neutrino components have a
negligible impact on the total spectrum. Compared to previous
solar neutrino analyses (\cite{Be7-1,Be7-2,Be7-3,BRX13}) an extended energy
region was used, including both $pp$ and $^{7}$Be neutrino contributions in
the same fit. In addition, the upper threshold of the fit is set above the
$^{11}$C end-point, which helps to constrain the resolution behaviour at the high end of the energy spectrum \cite{Brx17}.

The analytical model used to describe the data is an improved version of
the one described in~\cite{BRX13} with the goal of enlarging the fitting energy range.
The principal changes concern the non-linearities of the energy scale
and the addition of the resolution parameter to take into account the low-energy region. 
The former parameters were first used in the $pp$-neutrino
flux analysis \cite{PP14}. The energy estimator $N_{p}$ used is the number of PMTs triggered in each event (window of 230~ns) and is 
normalized to 2000 PMTs.

In the $pp$ neutrino analysis (408 days of data) non-normalized energy variables were used, but 
normalization was introduced as the number of live PMTs began to drop significantly.
In order to correct for the non-statistical fluctuations in the data from rebinning an intrinsically integer variable $N_{p}$ 
a correction at each bin was applied, calculated on the basis of the known number of functioning PMTs at each moment.
The model is discussed in \cite{Brx17}, and more detail will be presented in a devoted paper \cite{BrxModel}.

The analytical model function has in total 15 free parameters. The free parameters describing the energy scale and resolution are 
the light yield and two resolution parameters: one takes into account the
spatial non-uniformity of the detector's response and is relevant
for the high-energy part of the spectrum and the other is responsible
for the intrinsic resolution of the scintillator and effectively takes
into account other contributions to the scintillation response width
at low energies. Other parameters describe the rates of dominant backgrounds, namely
$^{14}$C (constrained to the value determined by analyzing an independent sample of $^{14}$C events selected
with low threshold, see \cite{PP14} for more detail), $^{85}$Kr, $^{210}$Bi,
$^{11}$C, $^{210}$Po peak, and external backgrounds (responses from the $^{208}$Tl and $^{214}$Bi $\gamma$-rays modelled with MC).
The $pp$ and $^{7}$Be interaction rates represent the solar neutrino parameters.
The remaining free parameters describe the position and width of
the $^{210}$Po $\alpha$-peak, and the starting point of the $^{11}$C $\beta^{+}$-
spectrum (corresponding to 2 annihilation gammas of 511 keV) as independent calibration doesn't provide the necessary precision to have them fixed or constrained.

The $pep$ and $^{8}$B solar neutrino contributions were kept fixed according to the Standard 
Solar Model (SSM) predictions and the uncertainty of the prediction contributed to the 
systematics as described in section {\ref{Syst}}. The minor contribution from external $^{40}$K $\gamma$-rays was fixed too.

Other parameters of the model are tuned either using MC modeling or
independent measurements and calibrations, for details see \cite{Brx17},~\cite{Brx_MC} and \cite{BrxModel}. 
They correspond to parameters describing the energy scale
nonlinearities: the ionization quenching parameter
, the contribution of the Cherenkov radiation
, the geometric correction to the energy scale
, the effective fraction of the single electron response under the threshold
; and an additional parameter in the resolution description
(quadratic with respect to the energy estimator). Special care
is taken to describe the pile-up events. 
The same approach is adopted as the one developed for the $pp$-neutrino analysis \cite{PP14}, 
where the synthetic pile-up is constructed by overlapping real events with randomly sampled data of the same time length.

\begin{figure}
\includegraphics[width=1\columnwidth]{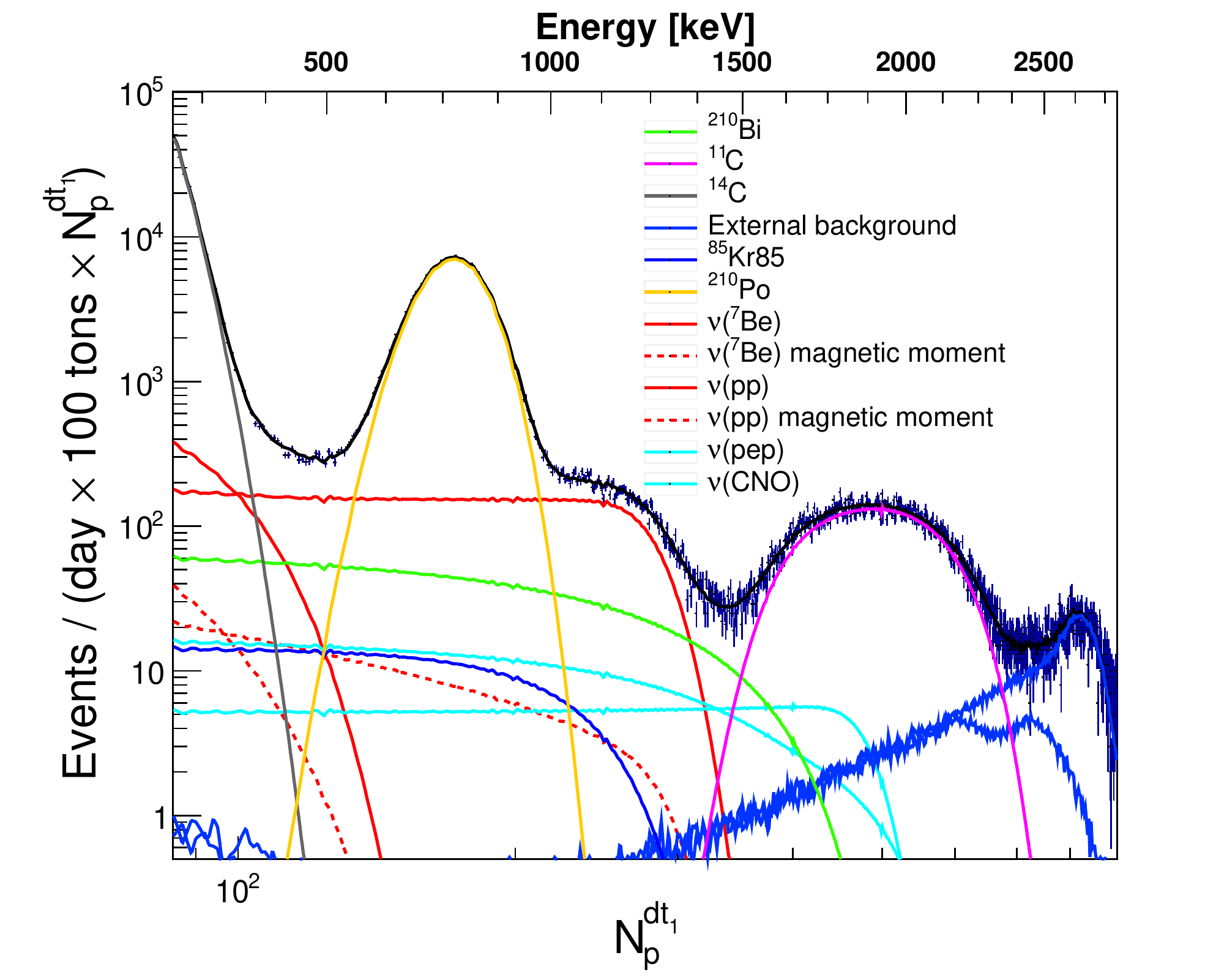}

\protect\caption{\label{Fig1}Spectral fit with the neutrino effective moment fixed at
$\mu_{\nu}^{eff}=2.8\times10^{-11}\mu_{B}$ (note the scale is double
logarithmic to underline the contributions at lower energies).}
\end{figure}

The $^{210}$Bi background and the CNO neutrino spectra are strongly anticorrelated
as they have similar spectral shapes. Their sum is constrained by 
the total number of events in the region between the $^{7}$Be Compton-like shoulder
and the $^{11}$C spectrum (see Fig.~\ref{Fig1}), which is mostly free
from other backgrounds. As the CNO contribution is masked 
by the larger $^{210}$Bi rate, the CNO neutrino rate is fixed to the 
SSM+MSW prediction without considering electromagnetic contribution. 
We used both high and low metallicity variants of the SSM, 
the difference in results was included in the systematics.
The electromagnetic term did not affect the fit results with respect 
to the CNO contribution as it was absorbed by the $^{210}$Bi component.

The likelihood profile as a function of $\mu_{\nu}^{eff}$ is obtained from the fit with the addition
of the electromagnetic component for $^{7}$Be and $pp$-neutrinos keeping
$\mu_{\nu}^{eff}$ fixed at each point. The electromagnetic contribution from
all other solar neutrino fluxes is negligible and is not considered
in the fit. Including the electromagnetic component described by (\ref{Formula:EM})
in the $pp$-neutrino cross-section leads to a decrease of
the $pp$-neutrino flux in the fit, compensating for the increase
in the total cross section. Another important correlation arises from
the presence of $^{85}$Kr in the fitting function. An increase in
the $^{7}$Be rate due to the electromagnetic
interactions is compensated for by a decrease in the $^{85}$Kr counting rate.
These two correlations in the fit decrease the overall sensitivity to the magnetic moment. 
The contribution from $^{85}$Kr
could be constrained from an independent measurement using a
delayed coincidence, but the combination of a very low branching ratio of
0.4\%, low tagging efficiency ($\sim$18\%), and a relatively
low $^{85}$Kr rate lead to very low statistics in the coincidence
branch~\cite{Brx17}. As a result, constraining $^{85}$Kr doesn't improve the sensitivity.
On the other hand, the correlation between the magnetic moment and the $pp$-neutrino flux
can be constrained by applying the results from radiochemical experiments, 
which are independent to the electromagnetic properties of neutrinos,
to the sum of the neutrino fluxes detected in Borexino.

The radiochemical constraints are based on the results from \cite{GaGe}.
The measured neutrino signal in gallium experiments expressed in Solar Neutrino Units (SNU) is:

$R={\displaystyle {\displaystyle \sum_{i}R_{i}^{Ga}=}\sum_{i}\Phi_{i}\intop}_{E_{th}}^{\infty}s_{i}^{\varodot}(E)P_{ee}(E)\sigma(E)dE=$

\begin{equation}
=\sum_{i}\Phi_{i}<\sigma_{i}^{\varodot}>=66.1\pm3.1~SNU,\label{eq:Gallium-1}
\end{equation}

where R is the total neutrino rate, $R_{i}$ is the contribution of the i-th solar
neutrino flux to the total rate, $\Phi_{i}$ is the neutrino flux from
i-th reaction, $s_{i}^{\varodot}(E)$ is the shape of the corresponding
neutrino spectrum in the Sun, $P_{ee}(E)$ is the electron neutrino survival
probability for neutrinos with energy E, and $\sigma(E)$ is the total cross-section
of the neutrino interaction with Ga which has a threshold of $E_{th}$=233
keV.

If applied to Borexino the radiochemical constraint takes the form:

\begin{equation}
\sum_{i}\frac{R_{i}^{Brx}}{R_{i}^{SSM}}R_{i}^{Ga}=(66.1\pm3.1\pm\delta_{R}\pm\delta_{FV})~SNU\label{eq:Radiochemical}
\end{equation}

where the expected gallium rates $R_{i}^{Ga}$ are estimated using new
survival probabilities of $P_{ee}$ based on recent values 
from \cite{OscPars} (therefore giving a new estimate for $<\sigma_{i}^{\varodot}>$), 
$\frac{R_{i}^{Brx}}{R_{i}^{SSM}}$ is the ratio of the corresponding
Borexino measured rate to its SSM prediction
within the MSW/LMA oscillation scenario. We used the same SSM predictions
for Borexino and the gallium experiments to avoid rescaling
the gallium expected rates. The total deviation from the measured
value should naturally include the additional theoretical error $\delta_{R}\simeq4\%$
from the uncertainty in estimating the single rates contributing
to the gallium experiments, and the uncertainty of the Borexino FV selection $\delta_{FV}\simeq1\%$.

Applying the radiochemical constraint (\ref{eq:Radiochemical}) to
the fit as an additional penalty term the analysis
of the likelihood profile gives a limit of $\mu_{\nu}^{eff}<2.6\cdot10^{-11}\mu_{B}$
at 90$\%$ C.L. for the effective magnetic moment of neutrinos using the
``standard'' fit conditions (230~ns time window energy variable, synthetic pile-up, high metallicity SSM and fixing the energy scale and resolution parameters).
Without radiochemical constraints the limit is weaker $\mu_{\nu}^{eff}<4.0\cdot10^{-11}\mu_{B}$
at 90$\%$ C.L. and is not used in the present analysis. An example of the spectral fit is presented in Fig.~\ref{Fig1}. 

\section{\label{Syst}Systematics study}

\begin{figure}
\includegraphics[width=1\columnwidth]{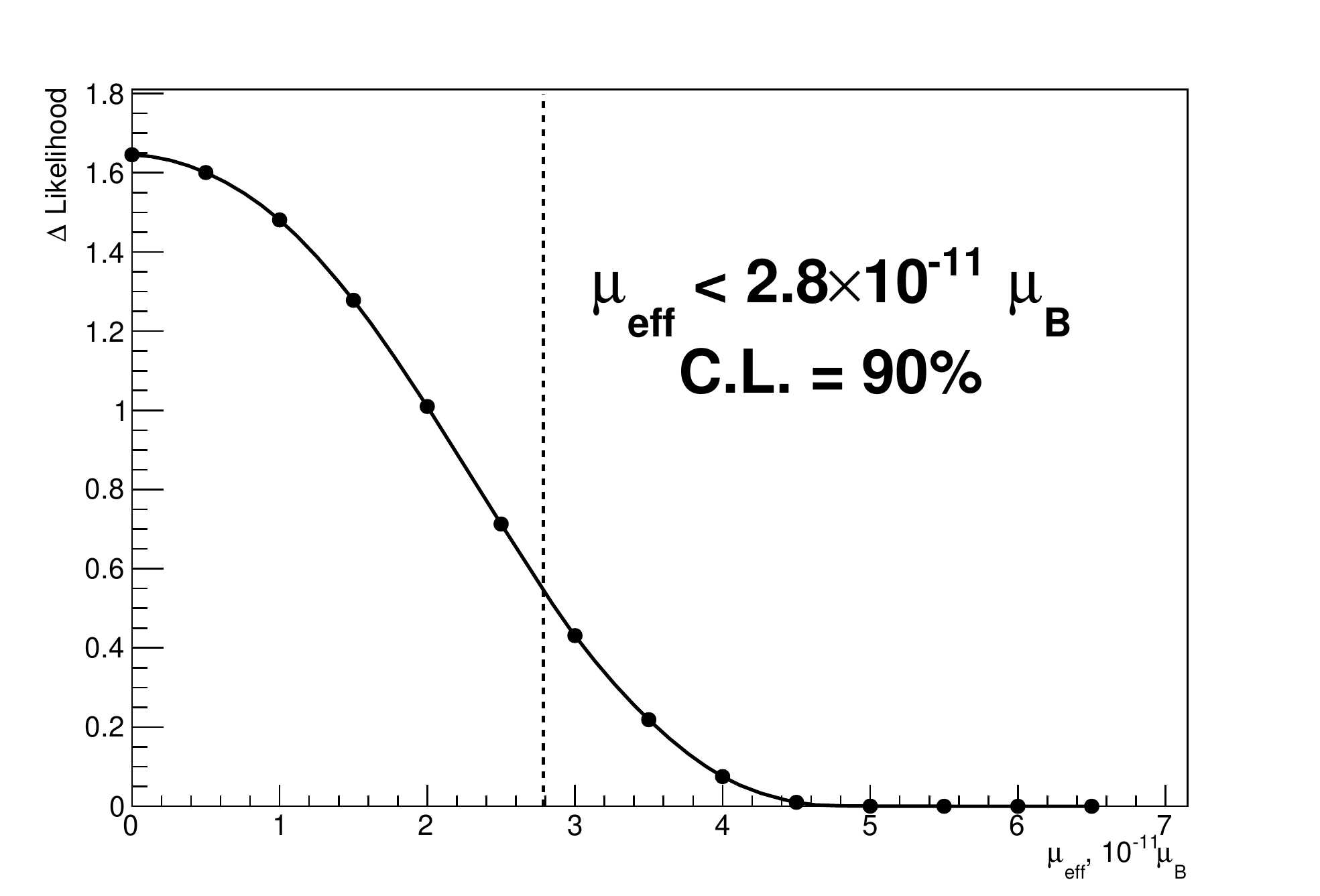}

\protect\caption{\label{Fig2}Resulting weighted likelihood profile used to estimate the limit on the neutrino magnetic moment. 
The profile doesn't follow the gaussian distribution as it is flatter initially and goes to zero faster than the normal distribution. 
The limit corresponds to 90\% of the total area under the curve. Note that unphysical values of $\mu_\nu^{eff}<0$ are not considered.}
\end{figure}

The systematics have been checked following the approach developed for
other Borexino data analyses~\cite{PP14,eDecay}. The main contributions
to the systematics comes from the difference in results depending on
the choice of energy estimator and the approach used for
the pile-up modelling. The energy estimators used in the analysis are the number
of PMTs triggered within a time window of 230 and 400~ns. The pile-up can be
reproduced by either convolving the model spectra with the data acquired from the
random trigger in the corresponding time window or by constructing a synthetic spectral component as described in~\cite{PP14}. 
Since the $pep$- and CNO- neutrino rates are fixed to the SSM predictions,
the different rates corresponding to high/low metallicity models are
also accounted for in the systematics. Further study included varying the fixed parameters within their expected errors.

The resulting likelihood profile
is the weighted sum of the individual profiles of each fit configuration. 
Initially, the same weights are used for the pile-up and SSM choice, assuming equal probabilities for all 4 possibilities. 
Further weights are assigned proportionally to the maximum likelihood of each profile, 
therefore taking into account the quality of the realization of the model with a given set of parameters. 
Accounting for the systematic uncertainties the limit on the effective neutrino
magnetic moment reduces to $\mu_{\nu}^{eff}$ $<$ 2.8$\cdot$10$^{-11}$
$\mu_{B}$ at 90\% C.L. The corresponding likelihood profile is shown in Fig.~\ref{Fig2}.

\section{Mass eigenstates basis}

Since neutrinos are a mixture of mass eigenstates the effective magnetic moment 
for neutrino-electron scattering is:

\begin{equation}
\mu_{eff}^{2}=\sum_{j}|\sum_{k}\mu_{kj}A_{k}(E_{\nu},L)|^{2}\label{mueff},
\end{equation}

where $\mu_{jk}$ is an element of the neutrino electromagnetic
moments matrix and $A_{k}(E_{\nu},L)$ is the amplitude of the k-mass state
at the point of scattering \cite{Bea99}. For the Majorana neutrino,
only the transition moments are non-zero, while the diagonal elements
of the matrix are equal to zero due to CPT-conservation. For the Dirac neutrino,
all matrix elements may have non-zero values. The effective magnetic
moment can be expanded both in terms of the mass eigenstates (this is more natural)
or the flavor eigenstates. Under the assumption that $\theta_{13}$
= 0, the form of the effective magnetic moment for the MSW oscillation
solution has been investigated in \cite{Jos02,Gri03}. The analysis
of Majorana transition neutrino magnetic moments taking into account
the non-zero value of the angle $\theta_{13}$ was first performed
in \cite{Can16}.

In the general case the expression for the effective magnetic moment in the mass eigenstate
basis will have a complex form consisting of interference terms $\propto\mu_{jk}\mu_{ik}$.
Without significant omissions the solar neutrinos arriving at the Earth
can be considered as an incoherent mixture of mass eigenstates \cite{Gri03,Dig99}.
In the case of Dirac neutrinos assuming that only diagonal magnetic
moments $\mu_{ii}$ are non-vanishing:

\begin{equation}
\mu_{eff}^{2}=P_{e1}^{3\nu}\mu_{11}^{2}+P_{e2}^{3\nu}\mu_{22}^{2}+P_{e3}^{3\nu}\mu_{33}^{2}\label{mueffDirac}
\end{equation}
where $P_{ei}^{3\nu}=|A_{i}(E,L)|^{2}$ is the probability of observing
the $i$-mass state at the scattering point for an initial electron flavor.

In the case of Majorana transition magnetic moments the effective
moment is:

\begin{equation}
\mu_{eff}^{2}=P_{e1}^{3\nu}(\mu_{12}^{2}+\mu_{13}^{2})+P_{e2}^{3\nu}(\mu_{21}^{2}+\mu_{23}^{2})+P_{e3}^{3\nu}(\mu_{31}^{2}+\mu_{32}^{2})\label{mueffMajorana}
\end{equation}

For the well known approximation of three- to two- neutrino
oscillation probabilities for solar neutrinos \cite{Gri03}: $P_{e1}^{3\nu}=\cos^{2}\theta_{13}P_{e1}^{2\nu}$,
$P_{e2}^{3\nu}=\cos^{2}\theta_{13}P_{e2}^{2\nu}$ and $P_{e3}^{3\nu}=\sin^{2}\theta_{13}$
-- one can get the effective magnetic moment expressed in well-established
oscillation parameters in the mass eigenstate basis. Equation (\ref{mueffDirac})
can be rewritten as:

\begin{equation}
\mu_{eff}^{2}=C_{13}^{2}P_{e1}^{2\nu}\mu_{11}^{2}+C_{13}^{2}P_{e2}^{2\nu}\mu_{22}^{2}+S_{13}^{2}\mu_{33}^{2}\label{mueffDirac2}
\end{equation}
where $C_{13}^{2}\equiv\cos^{2}\theta_{13}$ and $S_{13}^{2}\equiv\sin^{2}\theta_{13}$,
and $P_{e1}^{2\nu}+P_{e2}^{2\nu}=1$. Similarly, assuming CPT-conservation
($\mu_{jk}=\mu_{kj}$) relation (\ref{mueffMajorana}) for the
transition moments can be rewritten as:

\begin{equation}
\mu_{eff}^{2}=C_{13}^{2}P_{e1}^{2\nu}\mu_{12}^{2}+(1-C_{13}^{2}P_{e2}^{2\nu})\mu_{13}^{2}+(1-C_{13}^{2}P_{e1}^{2\nu})\mu_{23}^{2}\label{mueffMajorana2}
\end{equation}

In general, $P_{e1}^{2\nu}$ and $P_{e2}^{2\nu}$ (and $P_{ee}^{2\nu}$)
depend on the neutrino energy and shape of the neutrino spectrum ($pp$,
$^{7}$Be, $^{8}B$ etc.), but in the energy region below 1 MeV, a dominant
contribution to the recoil-electron spectrum comes from $pp$, $^{7}$Be,
and CNO neutrinos, for which dependence of $P_{e1}^{2\nu}$, $P_{e2}^{2\nu}$ and
$P_{ee}^{2\nu}$ on energy is weak and the probabilities can be assumed constant. 
Under such assumptions,
since $\mu_{eff}^{2}$ is the sum of positively defined quantities,
one can constrain any term in (\ref{mueffDirac2}) and (\ref{mueffMajorana2}).
By using the most probable values of $P_{ee}^{2\nu}$, $\theta_{13}$
and $\theta_{23}$ \cite{PDG16} one can obtain the following limits
from the relation $\mu_{eff}\leq2.8\times10^{-11}\mu_{B}$:

\begin{equation}
|\mu_{11}|\leq3.4;~~|\mu_{22}|\leq5.1;~~|\mu_{33}|\leq18.7;\label{limDirac}
\end{equation}
\begin{equation}
|\mu_{12}|\leq2.8;~~|\mu_{13}|\leq3.4;~~|\mu_{23}|\leq5.0;\label{limMajorana}
\end{equation}

all measured in units of $10^{-11}\mu_{B}$ and for 90$\%$
C.L..

\section{Limits on magnetic moments of the neutrino flavor states}

The effective magnetic moment for the LMA-MSW solution is (assuming the survival probability of $pp$ and $^{7}$Be solar neutrinos is the same):

\begin{equation}
\mu_{eff}^{2}=P^{3\nu}\mu_{e}^{2}+(1-P^{3\nu})(\cos^{2}\theta_{23}\cdot\mu_{\nu}^{2}+\sin^{2}\theta_{23}\cdot\mu_{\tau}^{2})\label{Formula:Meff-1},
\end{equation}

where $P^{3\nu}=\sin^{4}\theta_{13}+\cos^{4}\theta_{13}P^{2\nu}$
is the probability that $\nu_{e}$ is detected in its original flavor
(survival probability), with $P^{2\nu}$ calculated in the ``standard''
2-neutrino scheme, $\theta_{13}$ and $\theta_{23}$ are the corresponding
mixing angles. Though $P^{2\nu}$ depends on $E_{\nu}$, the difference
between $P^{2\nu}(400)=0.57$ for a neutrino energy close to the $pp$-neutrino spectrum
end-point of 420 keV (only a small fraction of the total $pp$-neutrino
spectrum close to the end-point contributes to the sensitive region
in our analysis) and $P^{2\nu}(862)=0.55$ for $^{7}$Be-neutrinos (higher
energy line) is negligible and we can reasonably assume them equal. 
Moreover, tests performed by ``turning on'' separately the
$pp$ and $^{7}$Be neutrino magnetic moments demonstrates that sensitivity
to the magnetic moment is dominated by the $^{7}$Be-neutrino contribution.
Therefore an estimate of $P^{2\nu}(400)=0.55$ is used in further calculations.

The limits on the flavor magnetic moment can be obtained from (\ref{Formula:Meff-1})
because individual contributions are positive. With $\mu_{\nu}^{eff}<2.8\cdot10^{-11}\mu_{B}$
and for $\sin^{2}\theta_{13}=0.0210\pm0.0011$ and $\sin^{2}\theta_{23}=0.51\pm0.04$
for normal hierarchy (or $\sin^{2}\theta_{23}=0.50\pm0.04$ for inverted
hierarchy)~\cite{PDG16} we obtain: $\mu_{e}<3.9\cdot10^{-11}\mu_{B}$,
$\mu_{\mu}<5.8\cdot10^{-11}\mu_{B}$ and $\mu_{\tau}<5.8\cdot10^{-11}\mu_{B}$,
all at 90\% C.L.

Because the mass hierarchy is still unknown, the values above were calculated for the
``unfortunate'' choice of hierarchy, providing conservative limits. 

\section{Conclusions}

New upper limits for the neutrino magnetic moments have been obtained
using 1291.5 days of data from the Borexino
detector. We searched for effects of the neutrino magnetic moments
by looking for distortions in the shape of the electron recoil spectrum.
A new model independent limit of $\mu_{\nu}^{eff}$<2.8$\cdot$10$^{-11}$
$\mu_{B}$ is obtained at 90\% C.L. including systematics. The
limit is free from uncertainties associated with predictions from the SSM neutrino flux and systematics from the detector's FV
 and is obtained by constraining the sum of the solar neutrino fluxes using the results from gallium experiments. 
The limit on the effective neutrino moment for solar neutrinos was used to set new limits on the magnetic moments for the neutrino flavor states and 
for the elements of the neutrino magnetic moments matrix for Dirac and Majorana neutrinos.

\section{Acknowledgements}

The Borexino program is made possible by funding from INFN (Italy),
NSF (USA), BMBF, DFG, HGF and MPG (Germany), RFBR (Grants 16-02-01026A,
15-02-02117A, 16-29-13014 ofi-m, 17-02-00305A ) and RSF (Grant 17-12-01009) (Russia), JINR Grant 17-202-01, and NCN
Poland (Grant UMO-2013/10/E/ST2/00180 ). We acknowledge the generous
hospitality and support of the Laboratory Nazionali del Gran Sasso
(Italy).

\end{document}